# Estimating Social Influence Using Latent Space Adjusted Approach in R

Ran Xu[a]

[a] Grado Department of Industrial and System Engineering, College of Engineering, Virginia Tech, Falls Church, Virginia, USA

## Abstract

Social influence, sometimes referred to as spillover or contagion, have been extensively studied in various empirical social network research. However, there are various estimation challenges in identifying social influence effects, as they are often entangled with other factors, such as homophily in the selection process, the individual's preference for the same social settings, etc. Methods currently available either do not solve these problems or require strong assumptions. Recent works by Xu (2018) and others show that a latent-space adjusted approach based on the latent space model has potential to disentangle the influence from other processes, and the simulation evidence shows the approach performs better than other state-of-the-art approaches in terms of recovering the true social influence effect when there is an unobserved trait co-determining influence and selection. In this paper we illustrate how latent-space adjusted approach accounts for bias in the estimation of the social influence effect, and demonstrate how this approach can be implemented to estimate various social influence models with an empirical example in R.

## 1. Introduction

Social influence, sometimes referred to as spillover or contagion, have long been central to the field of social science (Asch, 1952; Erbring and Young, 1979; Bandura, 1986). It is defined as the propensity for the behavior of an individual to vary along with the prevalence of that behavior in some reference group (Manski, 1993), such as one's social contacts. With the availability of social network data, social influence has received much attention and has been widely used to study various phenomenon such as the spread of health behavior (e.g. obesity and smoking) (Christakis et al., 2007, 2008), psychological states (Cacioppo et al., 2009; German et al., 2012), professional practices (Frank et al., 2004) and information diffusion (Valente, 1995, 1996).

However, social influence effects are usually difficult to identify, especially from observational network data, as it is difficult to separate the effect of influence from other processes that operate at the same time. That is, when we observe that people who are close to each other in terms of network distance (e.g. have a direct network tie) tend to be similar in some salient individual behavior and psychological states, it is difficult to identify the underlying mechanism that generates these patterns. It could be influence and contagion (Friedkin, 1999, 2001; Oetting and Donnermeyer, 1998) whereby actors assimilate the behavior of their network members; or selection mechanisms, more specifically homophily (McPherson and Smith-Lovin, 1987; McPherson et al., 2001), where actors seek to interact with similar others; or it could be due to common social-environmental factors where people with previous similarities can select themselves into the same social setting (e.g. school or social club), and actual friendship formation just reflects the opportunities of meeting in this social setting (Feld, 1981, 1982; Kalmijn & Flap, 2001).[1]

The entanglement between these different mechanisms unavoidably induces bias when we estimate the influence effect (Shalizi & Thomas, 2011). There are some current techniques that attempt to reduce the bias in estimating social influence effects, such as instrumental variable (IV) methods (Bramoullé et al., 2009), propensity score methods (Aral et al., 2009) and stochastic actor-oriented models (SAOM) (Snijder et al., 2010). Although each potentially leverages extra information in the data to reduce bias, none can claim to eliminate all sources of bias.

Recent works by Xu (2018) and others show that a latent-space adjusted approach based on the latent space model (Hoff, Raftery and Handcock, 2002) has potential to disentangle the influence from the other processes, and the simulation evidence shows the model performs well in terms of recovering the true influence effect. In this paper we illustrate how latent-space adjusted approach accounts for bias in the estimation of the social influence effect, and demonstrate how it can be applied to estimate various influence models with an empirical example in R. In the following sections, I will start by framing the bias in the estimation of social influence effect as an omitted variable bias problem. Then I will formally introduce the latent-space adjusted approach and how it can account for bias in the estimation of social influence effect. Finally I will demonstrate how to use the proposed approach to estimate social influence using the dynamic linear-in-mean influence model and the stochastic actor-oriented model (SAOM) with an empirical example in R.

## 2. Identification of Social Influence as An Omitted Variable Bias Problem

The similarity of the behavior, states, and characteristics of two individuals with a social tie can be caused by three primary mechanisms, namely social influence, homophilous selection or common social or environmental factors (Vanderweele & An, 2013). While it is possible to rule out some mechanisms through random assignment of treatment or networks in experiments, the entanglement between these different mechanisms makes it difficult to identify social influence effect from observational data. The difficulty of identification caused by entanglement between contagion effects and common social-environmental factors can be easily framed as an omitted variable bias problem (e.g. ignoring the group or environment individuals belong to when estimating the influence model). What is less obvious is that the dilemma caused by entanglement between the influence and homophilous selection can essentially be framed as an omitted variable bias problem as well. As pointed out by Steglich (2010), one of the important concerns of SAOM is the "possibility that there may be non-observed variables co-determining the probabilities of change in network and/or behavior". Shalizi and Thomas (2011) have shown that when there is an unobserved trait that co-determines both influence and selection in network data, social influence effects are generally unidentifiable, mainly due to the fact that social influence and homophily (selection) are generically confounded through this unobserved trait.

---

[1] There are also structural constraints such as transitivity, preferential attachment etc. which could cause people to become friends. However these mechanisms in themselves do not entangle with influence (e.g. one befriends with another having high popularity but different behavior). In these cases another mechanism must be present to induce similarity between these friends (e.g. selection of common friends based on similarity in attributes), and thus the entanglement goes back to the original three mechanisms, namely influence, selection based on homophily, and social-environmental factors.

To give an example, assuming that adolescent's delinquency behavior delinquency$_{it}$ is the outcome of interest, and it is a function of his/her previous delinquency behavior delinquency$_{it-1}$, his/her friend j's previous delinquency behavior delinquency$_{jt-1}$ (i.e. social influence), and an unobserved risk-taking tendency (arrow D in Figure 1). At the same time, when there is homophilous selection based on this unobserved risk-taking tendency in the networks, such that adolescents with similar level of risk-taking tendency are more likely to be friends (arrow A in Figure 1). As a result, person j's delinquency behavior, which is a function of person j's risk-taking tendency (arrow B$_j$), will be correlated with person i's risk-taking tendency through homophilous selection (arrow C in Figure 1). However, as the risk-taking tendency is unobserved, this violates the key assumption of most of the estimation methods (i.e. omitted variable should not correlate with the independent variables) such that the estimates of social influence will be biased and inconsistent.

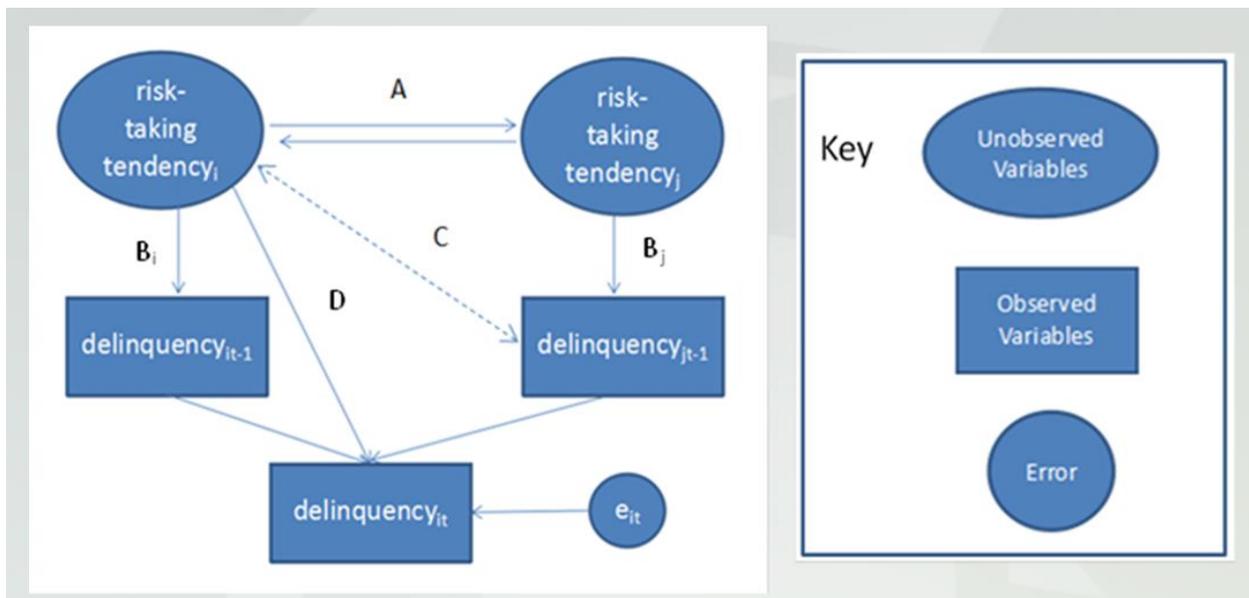

Figure 1: Omitted variable bias

## 3. Latent-space adjusted approach

Xu (2018) recently proposed a latent space adjusted approach that has potentials to correctly estimate the social influence effect when there is an unobserved variable that co-determines the influence and the selection process. Specifically, we represent a behavioral (influence) model can as

$$Y_{it} = f(Z_{ij}, Y_j, X_i, c_i) \qquad (1)$$

where the behavior of person *i* at time t is a function of the behavior of network members $Y_j$, other variables X specific to person i, network relations Z and unobserved variable $c_i$.[2] For example, adolescents' alcohol use ($Y_{it}$) can be a function of their previous alcohol use ($Y_{it-1}$), their close friends' alcohol use ($Y_{jt-1}$), their own cigarette use ($X_{it-1}$) and some latent disposition for substance abuse ($c_i$). The selection model can be represented as

$$P(Z_{ijt} = 1) = g(X_{ij}, D(c_i, c_j)) \qquad (2)$$

Where the probability that person i and person j has a network tie at time t is a function of individual and dyadic level observed variables $X_{ij}$, and a distance function of the unobserved variable c between i and j such that i and j are more likely to have tie when they are close to each other in terms of c. For example, the probability that adolescent i and j has a tie at time t ($Z_{ijt}$) can be a function of the absolute value of difference between their previous cigarette use $|X_{it-1}-X_{jt-1}|$ (observed homophily) and the absolute value of difference between latent disposition for substance abuse $|c_i-c_j|$ (latent homophily), where i and j are more likely to have tie when they are similar to each other in terms of cigarette use (X) or latent disposition for substance abuse (c).

Ideally if there is any information about this unobserved trait from the selection process in (2), it can be extracted and used in the estimation of the influence model in (1), and this will reduce the bias in estimating the contagion effects. However, the estimation of most selection models are based on observed variables and thus do not attend to those factors that are not observed. Xu (2018) builds on the theoretical logic of latent space models as applied to social-network data (Hoff et al., 2002). Latent space models assume that each individual has a "latent position" that lies in an unobserved n-dimensional social space, and the probability of interaction between any two actors depends on the latent positions of these two actors. Specifically, they take a logistic form and specify the selection model as

$$logodds(Z_{ij} = 1 | c_i, c_j, x_{ij}, \alpha, \beta) = \alpha + \beta' x_{ij} - |c_i - c_j| \qquad (3)$$

Here, $Z_{ij}$ indicates whether there is an interaction from i to j, $x_{ij}$ is a vector of observed covariates (at dyadic level or node level), c indicates the latent social position of i and j, and $|c_i - c_j|$ represents the Euclidean distance between i and j's latent position (it could also be replaced by other distance functions). A smaller distance between i and j's latent position indicates a larger probability of having a tie. And these latent social positions can be regarded as determinants of interactions that have not been accounted for by the observed variables in the selection process. The parameters α and β are estimated using either Maximum-Likelihood Estimation (MLE) or Markov Chain Monte Carlo (MCMC) methods, and the latent position c can be estimated by Minimum Kullback-Leibler (MKL) estimates (Shortreed et al., 2006).

It is not difficult to see that the latent space model in (3) is very similar to the selection process as we defined in (2), except that c represents the latent position in the latent space model, while c represents individual's unobserved trait in (2).[3] For any pair of i and j, a smaller distance between the latent social position or unobserved trait will result in a higher likelihood to have a network tie, and vice versa. Therefore, when two individuals are close to each other in terms of the unobserved trait, they are more likely to have a network tie and they should also be close to each other in terms of the latent social positions.

---
[2] Here we assume c is time invariant but the assumption can be relaxed.
[3] Here we only choose one-dimensional latent social positions to mimic the unobserved trait that drives the homophily in the selection process. The arguments can easily be extended to multi-dimensional latent positions.

Furthermore, if these latent positions from the latent space model are estimated accurately enough, the estimates of these latent positions can be used as proxies for the unobserved trait that determines the homophily in the selection process. In fact, for two one dimensional variables X and Y, if the distance correlation (e.g. correlation between $|X_i-X_j|$ and $|Y_i-Y_j|$) is 1, then Y can be written as a linear function of X: Y=a+bX (Szekely et al., 2007), which means the correlation between the two variables are either 1 or -1. Thus the estimated latent social positions from the latent space model can be included as a proxy for the unobserved trait when estimating an influence model, and this will in-principle reduce the bias in estimation of contagion effects that are due to the omitted variable problem (Wooldridge, 2010). For example, to model adolescents' delinquency behavior, we can first use a latent space model to model the friendship network of adolescents and acquire an estimated "latent social position" for each individual, and then use these estimates as proxies for the unobserved risk-taking tendency in the influence model, and thus achieve a better estimation of the true contagion effects. If the social network data is longitudinal, estimated latent social positions from different time points can be included in the influence model as separate covariates to better approximate the unobserved trait.

Shalizi and McFowland (2018) show that if the network grows according to a continuous latent space model, then latent homophilous attributes can be consistently estimated, and controlling for these latent attributes allows for unbiased and consistent estimation of social-influence effects in additive influence models. Simulation evidence from Xu (2018) show that when there is a time invariant unobserved variable that co-determines selection and influence, the estimated latent social positions can be good proxies for the unobserved variable, and the latent space adjusted approach outperforms other state-of-art estimation approaches in recovering the true social influence effect in a dynamic linear-in-mean influence model. The results are robust to the inclusion of additional covariates, structural properties (e.g. transitivity) in networks, different scaling of the latent space model, or even misspecifications (Xu, 2018).

Finally, there are a couple things to note: (1) for estimated latent positions from the latent space model to better approximate the unobserved trait, we need to control for other mechanisms that are likely to drive the selection process in the latent space model, such as homophily based on the observed variables, transitivity, alter and ego effect etc. (2) In principle this method can apply to any functional form of the influence model (e.g. stochastic actor-oriented models), as essentially this approach just adds additional covariates to approximate the unobserved trait. (3) As the scale and the actual position of the estimated latent social positions are essentially arbitrary (Hoff et al., 2002), the actual value of the latent social position might be very different from the actual value of the unobserved trait that codetermines influence and selection. However, as long as the estimated latent social positions are highly correlated with the unobserved trait (actors who are close to each other on the latent social positions are also close to each other in terms of the unobserved trait), the social influence effects can still be consistently estimated. (4) This approach specifically works for scenarios where there are unobserved traits that co-determine influence and selection (homophily).[4] It does not improve the estimation of the social influence when unobserved traits only present in one process but not the other.

---

[4] In principle this approach could also account for unobserved social-environmental factors that drives the influence and selection. There could also be multiple unobserved traits, in this case we use the latent social position to represent

## 4. An Empirical Example in R

In this section we present an empirical example illustrating how to use latent space adjusted approach to estimate the social influence effect in R 3.5.2. The data comes from the social network data collected in Teenage Friends and Lifestyle Study data set (Michell 2000, Pearson and West 2003). Friendship network data and substance use were recorded for a cohort of 50 female pupils in a school in the West of Scotland. The panel data were recorded over a three year period starting in 1995, when the pupils were aged 13, and ending in 1997. The friendship networks were formed by allowing the pupils to name up to twelve best friends. Pupils were also asked about substance use and adolescent behavior associated with, for instance, lifestyle, sporting behavior and tobacco, alcohol and cannabis consumption. The question on sporting activity asked if the pupil regularly took part in any sport, or go training for sport, out of school (e.g. football, gymnastics, skating, mountain biking). The school was representative of others in the region in terms of social class composition (Pearson and West 2003). The dataset is available at: https://www.stats.ox.ac.uk/~snijders/siena/s50_data.htm

First we install and load all the packages needed in R. latentnet is the package we use to estimate the latent space model.

```
> library(RSiena)
> library(sna)
> library(statnet)
> library(latentnet)
```

The network data comes with the RSiena package. We just need to load the attribute data into the current session, and create network objects over 3 time points:

```
##Load girls' attributes on smoking, drug use, sport and alcohol use
> s50s<-read.table("s50-smoke.dat",header=FALSE)
> s50d<-read.table("s50-drugs.dat",header=FALSE)
> s50sp<-read.table("s50-sport.dat",header=FALSE)
> s50a<-read.table("s50-alcohol.dat", header=FALSE)

## Create network object with attributes for each time point
> g1<-network(s501,directed=TRUE)
> g1%v%"a" <- s50a[,1]
> g1%v%"s" <- s50s[,1]
> g1%v%"sp" <- s50sp[,1]
> g1%v%"d" <- s50d[,1]

> g2<-network(s502,directed=TRUE)
> g2%v%"a" <- s50a[,2]
> g2%v%"s" <- s50s[,2]
```

---

a common latent construct of these unobserved traits.

```
> g2%v%"sp" <- s50sp[,2]
> g2%v%"d" <- s50d[,2]

> g3<-network(s503,directed=TRUE)
> g3%v%"a" <- s50a[,3]
> g3%v%"s" <- s50s[,3]
> g3%v%"sp" <- s50sp[,3]
> g3%v%"d" <- s50d[,3]
```

We can plot each network and see how it changes over time. Figure 2 shows how these girls' friendship network change from 1995 to 1997. The network graph shows that there is considerable network changes over time, and distinct components/clusters emerge over time.

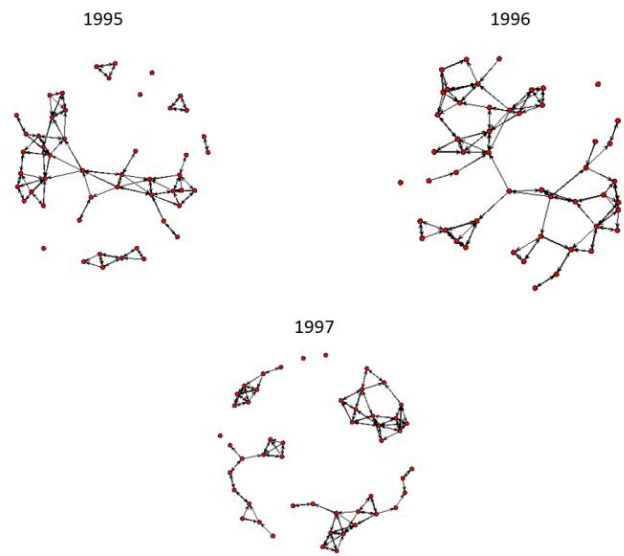

Figure 2. Girls' friendship network from 1995 to 1997.

Our primary research question is to study whether these girls influence each other's alcohol use. Here we demonstrate how to use latent space adjusted approach to estimate a dynamic linear-in-mean model (Friedkin & Johnsen, 1990) and a stochastic actor-oriented model (Snijder et al., 2010) in R. We start by estimating latent space models to extract the estimated latent positions. Specifically, we estimate two latent space models based on networks in 1995 and 1996 with one dimensional latent space, while controlling for homophily based on observed variables such as alcohol, smoking, drug use and sport:

```
> m1<-ergmm(g1 ~ euclidean(d = 1)+absdiff("a")+absdiff("s")+absdiff("sp")+absdiff("d"),control=ergmm.control(sample.size=5000,burnin=20000,interval=10,Z.delta=5))

> m2<-ergmm(g2 ~ euclidean(d = 1)+absdiff("a")+absdiff("s")+absdiff("sp")+absdiff("d"),control=ergmm.control(sample.size=5000,burnin=20000,interval=10,Z.delta=5))
```

Once we have estimated the latent space models, we can extract latent social positions and add them as additional covariates when estimating the influence model. First we estimate a dynamic linear-in-mean influence model, which can be represented as (Friedkin & Johnsen, 1990):

$$Y_{it} = \beta_0 + \beta_1 Y_{it-1} + \beta_2 \frac{\sum Z_{ijt-1} Y_{jt-1}}{\sum Z_{ijt-1}} + \beta_3 X_{it} + e_{it} \quad (4)$$

where $Y_{it}$ is behavior of i at time t, $Y_{it-1}$ is the previous behavior of $i$, $Z_{ijt-1}$ is a dummy variable indicating if there is a link from $i$ to $j$ at time t-1, i.e. 1 if yes and 0 otherwise, and $\frac{\sum Z_{ijt-1} Y_{jt-1}}{\sum Z_{ijt-1}}$ represents the weighted average behavior among the network neighbors of $i$, which is the exposure term of interest, and $X_{it}$ represents other concurrent variables that might affect the behavioral outcome Y.

To estimate the dynamic linear-in-mean influence model we first need to construct the dataset in R:

```
## create weighted average alcohol use of one's network neighbors
> E<-matrix(0,50,3)
for (i in 1:50)
{
if (sum(s50l[i,])!=0)
E[i,1]<-(s50l[i,]%*%s50a[,1])/sum(s50l[i,])
if (sum(s502[i,])!=0)
E[i,2]<-(s502[i,]%*%s50a[,2])/sum(s502[i,])
if (sum(s503[i,])!=0)
E[i,3]<-(s503[i,]%*%s50a[,3])/sum(s503[i,])
}
## create dataset to estimate the dynamic linear-in-mean influence model
> alcohol<-c(s50a[,3],s50a[,2])
> lag_alc<-c(s50a[,2],s50a[,1])
> expo<-c(E[,2],E[,1])
> drug<-c(s50d[,3],s50d[,2])
> smoke<-c(s50s[,3],s50s[,2])
> sport<-c(s50sp[,3],s50sp[,2])
> latent_pos2<-rep(m2$mkl$Z,2)
> latent_pos1<-rep(m1$mkl$Z,2)
> infl<-data.frame(cbind(alcohol,lag_alc,expo,drug,smoke,sport,
latent_pos1,latent_pos2,rep(c(1:50),2),rep(c(1:2),each=50)))
> head(infl)

  alcohol lag_alc    expo drug smoke sport latent_pos1 latent_pos2 V9 V10
1       3       1 4.333333    1     1     1   -5.997364   -8.472008  1   1
```

```
2       2       2 4.000000   3   3   1   -7.324663   -2.941830   2   1
3       3       3 2.500000   1   1   1    6.734962    9.064313   3   1
4       2       3 3.000000   1   1   1    6.734962    9.197778   4   1
5       4       3 3.500000   3   1   2    1.945568    7.413702   5   1
6       4       4 5.000000   1   3   2   18.585402    1.648355   6   1
```

To look at the correlation table between estimated latent positions and observed variables:

```
> cor(infl[,1:8])
```

```
            alcohol  lag_alc   expo    drug   smoke   sport  latent_pos1 latent_pos2
alcohol      1.0000   0.699   0.458   0.455   0.386  -0.092    -0.387      -0.317
lag_alc      0.6992   1.000   0.461   0.455   0.465  -0.165    -0.403      -0.364
expo         0.4585   0.461   1.000   0.348   0.416  -0.221    -0.550      -0.241
drug         0.4553   0.455   0.348   1.000   0.592  -0.382    -0.283      -0.453
smoke        0.3863   0.465   0.416   0.592   1.000  -0.224    -0.340      -0.463
sport       -0.0922  -0.165  -0.221  -0.382  -0.224   1.000     0.145       0.162
latent_pos1 -0.3872  -0.403  -0.550  -0.283  -0.340   0.145     1.000       0.150
latent_pos2 -0.3173  -0.364  -0.241  -0.453  -0.463   0.162     0.150       1.000
```

From the correlation table we can observe strong network autocorrelation - one's alcohol use (alcohol), previous alcohol use (lag_alc), and friends' alcohol use (expo) are all highly correlated with each other. Furthermore, the estimated latent social positions from 1995 and 1996 (latent_pos1 and latent_pos2) have sizable correlations with both one's alcohol use and one's friends' alcohol use. As the calculation of latent social positions is conditioned on homophily based on observed variables such as alcohol, drug, smoking and sport, the results suggest that there might be some unobserved variables (e.g. latent propensity for substance abuse) that drive both influence and selection.

To estimate the dynamic linear-in-mean influence model, we first estimate an influence model without latent social positions, then we estimate another one with latent social positions included as additional covariates:

```
> summary(lm(alcohol~lag_alc+expo+smoke+sport+drug,data=infl))

Call:
lm(formula = alcohol ~ lag_alc + expo + smoke + sport + drug,
    data = infl)

Residuals:
    Min      1Q  Median      3Q     Max
-2.2382 -0.4876  0.0384  0.4935  1.6371

Coefficients:
            Estimate Std. Error t value Pr(>|t|)
(Intercept)  0.47833    0.40080   1.193   0.2357
lag_alc      0.53760    0.08160   6.588 2.54e-09 ***
expo         0.15298    0.07516   2.035   0.0446 *
smoke       -0.05760    0.11602  -0.496   0.6207
sport        0.23565    0.16698   1.411   0.1615
drug         0.26057    0.11873   2.195   0.0307 *
---
Signif. codes:  0 '***' 0.001 '**' 0.01 '*' 0.05 '.' 0.1 ' ' 1

Residual standard error: 0.7655 on 94 degrees of freedom
```

```
Multiple R-squared:  0.5398,	Adjusted R-squared:  0.5154
F-statistic: 22.06 on 5 and 94 DF,  p-value: 1.473e-14

> summary(lm(alcohol~lag_alc+expo+smoke+sport+drug+latent_pos1+latent_pos2,data=infl))

Call:
lm(formula = alcohol ~ lag_alc + expo + smoke + sport + drug +
    latent_pos1 + latent_pos2, data = infl)

Residuals:
    Min      1Q  Median      3Q     Max
-2.2031 -0.5060  0.1155  0.5177  1.6341

Coefficients:
             Estimate Std. Error t value Pr(>|t|)
(Intercept)  0.625653   0.453098   1.381   0.1707
lag_alc      0.525024   0.084136   6.240 1.32e-08 ***
expo         0.128865   0.083382   1.545   0.1257
smoke       -0.071843   0.120567  -0.596   0.5527
sport        0.235112   0.168289   1.397   0.1658
drug         0.251790   0.122337   2.058   0.0424 *
latent_pos1 -0.007709   0.011007  -0.700   0.4854
latent_pos2 -0.004525   0.014706  -0.308   0.7590
---
Signif. codes:  0 '***' 0.001 '**' 0.01 '*' 0.05 '.' 0.1 ' ' 1

Residual standard error: 0.7715 on 92 degrees of freedom
Multiple R-squared:  0.5426,	Adjusted R-squared:  0.5078
F-statistic: 15.59 on 7 and 92 DF,  p-value: 2.541e-13
```

Results show that conditioning on previous alcohol use and other observed covariates, the social influence effect on alcohol use is significant (coef=.152, se=.075, p=.045) – that is, if girls' friends use more alcohol, they will use more alcohol themselves. However, if we include latent positions as additional covariates, the social influence effect (coef=.129, se=.083, p=.126) is no longer significant.[5] This suggests that there are likely to be unobserved variables that drive both girls' alcohol use and selection, and ignoring them will lead to ~18% overestimation of social influence effect in this case, which leads to erroneous statistical inference.

Next we estimate a Stochastic Actor-oriented Model (SAOM) using RSiena testing if there is any social influence effect on girls' alcohol use. SAOM is a class of simulation based statistical models that can model the behavioral and network change simultaneously (Snijder et al., 2010). We start by constructing a dataset that can be used by SAOM models for estimation:

```
## create data structure that can be used to estimate SIENA
> friend.data.w1 <- s501
> friend.data.w2 <- s502
> friend.data.w3 <- s503
> drink <- s50a
> smoke <- s50s
> drug  <- s50d
> sport <- s50sp
> friendship <- sienaDependent( array( c( friend.data.w1, friend.data.w2,
+                                          friend.data.w3 ),
+                                 dim = c( 50, 50, 3 ) ) )
> drinkingbeh <- sienaDependent( drink, type = "behavior" )
> smokingbeh <- varCovar( as.matrix(smoke))
> drugbeh <- varCovar( as.matrix(drug))
> sportbeh <- varCovar( as.matrix(sport))
> lat1<-coCovar(as.vector(m1$mkl$Z)) ## latent position from 1995
```

---

[5] Latent space model uses a MCMC estimation and thus the results will be slightly different each time. It is suggested to estimate latent space model with longer burn-in, larger sample size, and over multiple times to acquire the final estimates (e.g. using mean or mode of the estimates).

```
> lat2<-coCovar(as.vector(m2$mkl$Z)) ## latent position from 1996
> myCoEvolutionData <- sienaDataCreate( friendship, drinkingbeh,smokingbeh,drugbeh,spo
rtbeh,lat1,lat2 )
```

To specify SAOM model , we type in the following codes. Specifically, in the selection part of the model we include structural effects such as reciprocity, transitivity, popularity, geometrically weighted degree, and homophily based on alcohol, drug use, smoking, sport and latent positions. In the behavioral part of the model we model girls' alcohol use as a function of linear and quadratic shape, average similarity effect (social influence), observed covariates such as drug use, smoking, sport, as well as latent positions as additional covariates:

```
> myCoEvolutionEff2 <- getEffects( myCoEvolutionData )
>
> effectsDocumentation(myCoEvolutionEff2)
>
> myCoEvolutionEff2 <- includeEffects( myCoEvolutionEff2, transTrip, cycle3,gwespFF,i
nPop,outPop)
> myCoEvolutionEff2 <- includeEffects( myCoEvolutionEff2, simX, interaction1 = "smokin
gbeh" )
> myCoEvolutionEff2 <- includeEffects( myCoEvolutionEff2, simX, interaction1 = "drugbe
h" )
> myCoEvolutionEff2 <- includeEffects( myCoEvolutionEff2, simX, interaction1 = "sportb
eh" )
> myCoEvolutionEff2 <- includeEffects(myCoEvolutionEff2,  simX,interaction1 = "drinkin
gbeh" )
> myCoEvolutionEff2 <- includeEffects(myCoEvolutionEff2,  simX,interaction1 = "lat1" )
> myCoEvolutionEff2 <- includeEffects(myCoEvolutionEff2,  simX,interaction1 = "lat2" )
> myCoEvolutionEff2 <- includeEffects( myCoEvolutionEff2,
+                                     name = "drinkingbeh",
+                                     avSim,
+                                     interaction1 = "friendship" )
> myCoEvolutionEff2 <- includeEffects( myCoEvolutionEff2,
+                                     name = "drinkingbeh", effFrom,
+                                     interaction1 = "smokingbeh")
> myCoEvolutionEff2 <- includeEffects( myCoEvolutionEff2,
+                                     name = "drinkingbeh", effFrom,
+                                     interaction1 = "drugbeh")
> myCoEvolutionEff2 <- includeEffects( myCoEvolutionEff2,
+                                     name = "drinkingbeh", effFrom,
+                                     interaction1 = "sportbeh")
> myCoEvolutionEff2 <- includeEffects( myCoEvolutionEff2,
+                                     name = "drinkingbeh", effFrom,
+                                     interaction1 = "lat1")
> myCoEvolutionEff2 <- includeEffects( myCoEvolutionEff2,
+                                     name = "drinkingbeh", effFrom,
+                                     interaction1 = "lat2")
```

To estimate the SAOM model we type:

```
> betterCoEvAlgorithm <- sienaAlgorithmCreate( projname = 's50CoEv_3',
+                                             diagonalize = 0.2, doubleAveraging = 0)
>
>
> (ans2 <- siena07( betterCoEvAlgorithm, data = myCoEvolutionData,
+                 effects = myCoEvolutionEff2))

Estimates, standard errors and convergence t-ratios

                                                  Estimate   Standard     Convergence
                                                             Error        t-ratio
Network Dynamics
   1. rate constant friendship rate (period 1)   6.7804   ( 1.9520   )    -0.0247
   2. rate constant friendship rate (period 2)   5.5804   ( 1.5074   )    -0.0446
   3. eval outdegree (density)                  -3.8226   ( 0.4710   )    -0.0268
   4. eval reciprocity                           2.2901   ( 0.4352   )    -0.0036
   5. eval transitive triplets                  -1.1221   ( 0.9042   )    -0.0228
   6. eval 3-cycles                              1.1517   ( 0.5529   )    -0.0201
```

```
  7. eval GWESP I -> K -> J (69)                      2.7667  ( 1.8913  )   -0.0204
  8. eval indegree - popularity                       0.1178  ( 0.1121  )   -0.0313
  9. eval outdegree - popularity                     -0.5368  ( 0.1570  )   -0.0286
 10. eval lat1 similarity                             0.6507  ( 0.5186  )   -0.0439
 11. eval lat2 similarity                             7.7015  ( 1.3888  )   -0.0198
 12. eval drinkingbeh similarity                      0.6110  ( 0.6676  )    0.0086
 13. eval smokingbeh similarity                       0.0755  ( 0.2577  )    0.0176
 14. eval drugbeh similarity                          0.8831  ( 0.5177  )   -0.0197
 15. eval sportbeh similarity                         0.2044  ( 0.1843  )    0.0627

Behavior Dynamics
 16. rate rate drinkingbeh (period 1)                 1.2506  ( 0.3943  )    0.0629
 17. rate rate drinkingbeh (period 2)                 1.7510  ( 0.5416  )    0.0174
 18. eval drinkingbeh linear shape                    0.3880  ( 0.1903  )   -0.0095
 19. eval drinkingbeh quadratic shape                -0.1304  ( 0.1459  )   -0.0447
 20. eval drinkingbeh average similarity              3.0265  ( 2.2662  )    0.0059
 21. eval drinkingbeh: effect from lat1              -0.0240  ( 0.0235  )    0.0612
 22. eval drinkingbeh: effect from lat2              -0.0169  ( 0.0324  )    0.0166
 23. eval drinkingbeh: effect from smokingbeh       -0.3243  ( 0.3157  )   -0.0706
 24. eval drinkingbeh: effect from drugbeh            0.0538  ( 0.2728  )   -0.0154
 25. eval drinkingbeh: effect from sportbeh           0.3266  ( 0.3720  )    0.0066

Overall maximum convergence ratio:    0.1721

Total of 3944 iteration steps.
```

Results show that there is strong homophily based on the latent positions in the selection process. And the estimate for average similarity (social influence) effect is 3.03, and the standard error is 2.27. Next we compare it with a SAOM model that excludes the latent positions in both selection and behavioral model:

```
Estimates, standard errors and convergence t-ratios

                                                    Estimate    Standard    Convergence
                                                                 Error        t-ratio
Network Dynamics
  1. rate constant friendship rate (period 1)        5.6744  ( 1.4262  )    0.0123
  2. rate constant friendship rate (period 2)        4.4861  ( 0.9524  )   -0.0206
  3. eval outdegree (density)                       -2.3732  ( 0.2822  )    0.0193
  4. eval reciprocity                                3.0429  ( 0.4632  )    0.0205
  5. eval transitive triplets                       -1.4128  ( 0.8951  )    0.0193
  6. eval 3-cycles                                   1.7027  ( 0.5465  )    0.0205
  7. eval GWESP I -> K -> J (69)                     3.6722  ( 1.7601  )    0.0104
  8. eval indegree - popularity                      0.0872  ( 0.1019  )   -0.0118
  9. eval outdegree - popularity                    -0.6361  ( 0.1700  )    0.0215
 10. eval drinkingbeh similarity                     1.2178  ( 0.7357  )    0.0277
 11. eval smokingbeh similarity                     -0.0006  ( 0.2812  )   -0.0166
 12. eval drugbeh similarity                         0.9889  ( 0.4224  )   -0.0231
 13. eval sportbeh similarity                        0.1628  ( 0.1859  )   -0.0149

Behavior Dynamics
 14. rate rate drinkingbeh (period 1)                1.2869  ( 0.3117  )    0.0219
 15. rate rate drinkingbeh (period 2)                1.7214  ( 0.4520  )    0.0173
 16. eval drinkingbeh linear shape                   0.3975  ( 0.1840  )   -0.0159
 17. eval drinkingbeh quadratic shape               -0.0542  ( 0.1209  )    0.0014
 18. eval drinkingbeh average similarity             4.0685  ( 2.0968  )    0.0147
 19. eval drinkingbeh: effect from smokingbeh      -0.2452  ( 0.3031  )    0.0476
 20. eval drinkingbeh: effect from drugbeh           0.0836  ( 0.2829  )    0.0277
 21. eval drinkingbeh: effect from sportbeh          0.3029  ( 0.3710  )   -0.0065

Overall maximum convergence ratio:    0.1545

Total of 3743 iteration steps.
```

The estimates for social influence effect is 4.07, with a standard error of 2.10. As a result, ignoring the latent position will likely lead to 34% overestimation of social influence effect in this case using the SAOM model.

## 5. Discussion and Conclusion

Social influence effects are generally difficult to identify, as influence processes are often entangled with other processes such as selection and environmental factors. Here we show that this entanglement/difficulty can essentially be framed as an omitted variable bias problem, and a latent space adjusted approach holds promise to correctly identify contagion effects in this case. And we demonstrate how to use latent space adjusted approach to estimate various social influence models with an empirical example in R. Results show that influence models ignoring the unobserved variables that drive both influence and selection are likely to overestimate the true social influence effect, while the latent space adjusted approach holds promise to correct that bias and serve as a more conservative test of the true social influence effect

Although the latent space adjusted approach proposed in this paper is flexible enough to be incorporated with any functional form of the influence model, and it holds much promise as an alternative approach to identify the social influence effect, there are also several limitations with this approach: (1) As previously mentioned, the latent space adjusted approach requires that the same unobserved traits occur in both influence and selection process. It can not account for the unobserved traits that are only present in one of the processes but not the other. (2) The choice of dimensions of latent social space in the latent space model is not clear. Although we choose one-dimensional latent social positions in all of the simulations and empirical examples, this needs not to be the case and there is no clear rule deciding how many dimensions users should use. (3) The computation of latent social position is very time consuming, and the computation time increases significantly with the increase of data or the number of dimensions in latent social position.

Nevertheless, we do believe that the latent space adjusted approach proposed here can provide a more plausible estimate of the true social influence effect, especially when the entanglement between influence and selection is of concern. Our study took a major step in clarifying the estimation challenges in the identification of social influence, providing a broad framework for a more plausible estimation of the social influence, and pointing to many future avenues of research.